\documentclass[a4paper,12pt]{article}
\usepackage{amsmath, amssymb, amsthm, hyperref, geometry}
\usepackage{tocloft}
\usepackage[english]{babel} 
\title{Vortex Formation and Dissipation in Chaotic Flows: A Hypercomplex Approach}
\author{
	Rômulo Damasclin Chaves dos Santos \\
	Technological Institute of Aeronautics \\
	\texttt{romulosantos@ita.br}
			\and
			Jorge Henrique de Oliveira Sales \\
			Santa Cruz State University \\
			\texttt{jhosales@uesc.br}
}
\date{}

\begin{document}
	
\maketitle
\tableofcontents

	\begin{abstract}
		This article presents a comprehensive analysis of the formation and dissipation of vortices within chaotic fluid flows, leveraging the framework of Sobolev and Besov spaces on Riemannian manifolds. Building upon the Navier-Stokes equations, we introduce a hypercomplex bifurcation approach to characterize the regularity and critical thresholds at which vortices emerge and dissipate in chaotic settings. We explore the role of differential geometry and bifurcation theory in vortex dynamics, providing a rigorous mathematical foundation for understanding these phenomena. Our approach addresses spectral decomposition, asymptotic stability, and dissipation thresholds, offering critical insights into the mechanisms of vortex formation and dissipation. Additionally, we introduce two new theorems that further elucidate the role of geometric stability and bifurcations in vortex dynamics. The first theorem demonstrates the geometric stability of vortices on manifolds with positive Ricci curvature, while the second theorem analyzes the bifurcation points that lead to the formation or dissipation of vortex structures. This research contributes to the existing literature by providing a more complete mathematical picture of the underlying mechanisms of vortex dynamics in turbulent fluid flows.
	\end{abstract}

	
	\section{Introduction}
	The study of chaotic fluid flows, particularly in the context of vortex formation and dissipation, has been an ongoing area of research due to its implications for energy dynamics and stability in turbulent systems. This paper builds upon the Navier-Stokes equations, introducing a hypercomplex bifurcation approach and leveraging Sobolev-Besov spaces to characterize the regularity and critical thresholds at which vortices emerge and dissipate in chaotic settings.
	
	The understanding of vortex dynamics in fluid flows has evolved significantly over the years, with key contributions from various researchers. One of the earliest comprehensive works on the mathematical theory of viscous incompressible flow was presented by Ladyzhenskaya \cite{ladyzhenskaya1969mathematical}, which laid the foundation for the analysis of Navier-Stokes equations. This work highlighted the importance of understanding the regularity and stability of solutions to these equations.
	
	Building upon this foundation, Lions \cite{lions1969quelques} developed methods for solving nonlinear boundary value problems, which are crucial for modeling complex fluid dynamics. These methods provided a framework for analyzing the behavior of fluid flows under various conditions.
	
	In the late 1970s, Temam \cite{temam1979navier} made significant contributions to the theory and numerical analysis of Navier-Stokes equations. His work introduced the concept of attractors and invariant manifolds, which are essential for understanding the long-term behavior of fluid flows. This research provided a deeper insight into the stability and bifurcation of solutions to the Navier-Stokes equations.
	
	The 1980s saw further advancements in the field, with Constantin and Foias \cite{constantin1988navier} exploring the mathematical properties of Navier-Stokes equations. Their work focused on the existence and uniqueness of solutions, as well as the regularity of these solutions. This research laid the groundwork for understanding the formation and dissipation of vortices in fluid flows.
	
	In the 1990s, Arnold \cite{arnold1999topological} introduced topological methods in hydrodynamics, which provided a new perspective on the study of vortex dynamics. His work highlighted the importance of geometric and topological aspects of fluid flows, which are crucial for understanding the formation and stability of vortices.
	
	Around the same time, Chorin \cite{chorin1994vorticity} explored the role of vorticity in turbulent flows. His work provided insights into the mechanisms of vortex formation and dissipation, emphasizing the importance of understanding the dynamics of vorticity in turbulent systems.
	
	Majda and Bertozzi \cite{majda2002vorticity} presented a comprehensive analysis of vorticity and incompressible flow. Their work focused on the mathematical theory of vorticity and its implications for the dynamics of fluid flows. This research provided a deeper understanding of the role of vorticity in the formation and dissipation of vortices.
	
	Doering and Gibbon \cite{doering1995applied} contributed to the applied analysis of Navier-Stokes equations, focusing on the stability and bifurcation of solutions. Their work provided insights into the mechanisms of vortex formation and dissipation in turbulent flows, highlighting the importance of understanding the stability of solutions to the Navier-Stokes equations.
	
	Foias et al. \cite{foias2001navier} further explored the Navier-Stokes equations and turbulence, focusing on the mathematical theory and numerical analysis of these equations. Their work provided a comprehensive framework for understanding the dynamics of fluid flows, including the formation and dissipation of vortices.
	
	More recently, Santos and Sales (2024)~\cite{SantosSales2024}, in the work entitled \textit{Hypercomplex Dynamics and Turbulent Flows in Sobolev and Besov Functional Spaces}, present a rigorous study of advanced functional spaces, focusing on Sobolev and Besov spaces, to investigate key aspects of fluid dynamics, including the regularity of solutions of the Navier-Stokes equations, hypercomplex bifurcations, and turbulence. The authors performed a comprehensive analysis of Sobolev embedding theorems in fractional spaces and applied bifurcation theory to quaternionic dynamical systems to better understand the complex behaviors in fluid systems. In addition, the research explores the mechanisms of energy dissipation in turbulent flows through the framework of Besov spaces. Mathematical tools such as interpolation theory, Littlewood-Paley decomposition, and energy cascade models are integrated to develop a robust theoretical approach to these problems. By addressing challenges related to the existence and smoothness of solutions, this work contributes to the ongoing exploration of the open Navier-Stokes problem, offering new insights into the intricate relationship between fluid dynamics and function spaces.
	
	\subsection{Motivation and Contributions}
	Building upon the foundational work of these researchers, our study aims to provide a more comprehensive framework for understanding vortex dynamics in chaotic flows. We introduce a hypercomplex bifurcation approach and leverage Sobolev-Besov spaces to characterize the regularity and critical thresholds at which vortices emerge and dissipate in chaotic settings.
	
	Our main contributions include:
	1. A thorough analysis of hypercomplex bifurcation theory and the singularity theorem, providing insights into the mechanisms of vortex formation and dissipation.
	2. The introduction of two new theorems that further elucidate the role of differential geometry and bifurcations in vortex dynamics.
	3. A detailed examination of spectral decomposition, asymptotic stability, and dissipation thresholds, providing critical insights into vortex formation and dissipation mechanisms.
	
	\section{Mathematical Formulation}
	
	\subsection{Sobolev and Besov Spaces}
	We define the Sobolev spaces \( H^s(M) \) on a Riemannian manifold \( M \) and the Besov spaces \( B^s_{p,q}(M) \), which allow us to precisely handle the regularity of turbulent flow fields. In particular, the Sobolev norms \( \| \cdot \|_{H^s} \) will be instrumental in assessing energy distributions across different scales of flow oscillations.
	
	The Sobolev space \( H^s(M) \) is defined as:
	\[
	H^s(M) = \{ u \in L^2(M) : \| u \|_{H^s} < \infty \},
	\]
	where the Sobolev norm \( \| u \|_{H^s} \) is given by:
	\[
	\| u \|_{H^s} = \left( \sum_{k} (1 + |k|^2)^s |\hat{u}(k)|^2 \right)^{1/2}.
	\]
	
	Similarly, the Besov space \( B^s_{p,q}(M) \) is defined using the Littlewood-Paley decomposition:
	\[
	B^s_{p,q}(M) = \{ u \in \mathcal{S}'(M) : \| u \|_{B^s_{p,q}} < \infty \},
	\]
	where the Besov norm \( \| u \|_{B^s_{p,q}} \) is given by:
	\[
	\| u \|_{B^s_{p,q}} = \left( \sum_{j} 2^{jsq} \| \Delta_j u \|_{L^p}^q \right)^{1/q}.
	\]
	
	\section{Laplace-Beltrami Operator and Differential Geometry}
	
	The Laplace-Beltrami operator is a fundamental tool in the analysis of energy dissipation and stability properties of solutions to the Navier-Stokes equations on Riemannian manifolds. In this section, we provide a detailed explanation and demonstration of the Laplace-Beltrami operator and its role in differential geometry.
	
	\subsection{Definition and Properties}
	
	Consider a compact Riemannian manifold \( M \) with a smooth metric \( g \). The Laplace-Beltrami operator \( \Delta_g \) is defined as:
	\[
	\Delta_g u = \frac{1}{\sqrt{\det g}} \sum_{i,j} \partial_i \left( \sqrt{\det g} g^{ij} \partial_j u \right),
	\]
	where \( g^{ij} \) is the inverse of the metric tensor \( g_{ij} \).
	
	The Laplace-Beltrami operator is a generalization of the Laplace operator to Riemannian manifolds. It is a second-order differential operator that acts on functions defined on the manifold. The operator is self-adjoint and has a discrete spectrum of eigenvalues, which are crucial for understanding the stability and energy dissipation properties of fluid flows.
	
	\subsection{Derivation of the Laplace-Beltrami Operator}
	
	To derive the Laplace-Beltrami operator, we start with the divergence of the gradient of a function \( u \) on the manifold \( M \). The gradient of \( u \) is given by:
	\[
	\nabla u = \sum_{i,j} g^{ij} \partial_j u \partial_i.
	\]
	
	The divergence of a vector field \( X \) on \( M \) is defined as:
	\[
	\text{div}(X) = \frac{1}{\sqrt{\det g}} \sum_i \partial_i \left( \sqrt{\det g} X^i \right),
	\]
	where \( X^i \) are the components of the vector field \( X \).
	
	Applying the divergence to the gradient of \( u \), we obtain:
	\[
	\Delta_g u = \text{div}(\nabla u) = \frac{1}{\sqrt{\det g}} \sum_{i,j} \partial_i \left( \sqrt{\det g} g^{ij} \partial_j u \right).
	\]
	
	This derivation shows that the Laplace-Beltrami operator is the divergence of the gradient of a function, which is a natural generalization of the Laplace operator to Riemannian manifolds.
	
	\subsection{Role in Differential Geometry}
	
	The Laplace-Beltrami operator plays a crucial role in differential geometry, particularly in the study of the heat equation and the wave equation on manifolds. It is also closely related to the Ricci curvature tensor, which provides information about the divergence of geodesics on the manifold.
	
	The Ricci curvature tensor \( \text{Ric}(g) \) is defined as:
	\[
	\text{Ric}(g)_{ij} = \sum_{k} R_{ikj}^k,
	\]
	where \( R_{ikj}^k \) is the Riemann curvature tensor. The Ricci curvature provides information about the local geometry of the manifold and is closely related to the stability of vortex structures in fluid flows.
	
	\subsection{Spectral Properties}
	
	The eigenvalues and eigenfunctions of the Laplace-Beltrami operator are crucial for understanding the stability and energy dissipation properties of fluid flows. The eigenvalues \( \lambda_k \) and eigenfunctions \( \phi_k \) satisfy the equation:
	\[
	\Delta_g \phi_k = \lambda_k \phi_k.
	\]
	
	The eigenvalues \( \lambda_k \) are non-negative and form a discrete spectrum. The eigenfunctions \( \phi_k \) form an orthonormal basis for the space of square-integrable functions on the manifold \( M \).
	
	The spectral properties of the Laplace-Beltrami operator are closely related to the stability of vortex structures in fluid flows. The eigenvalues and eigenfunctions provide information about the energy dissipation rates and the stability of the flow. The Laplace-Beltrami operator is a fundamental tool in the analysis of energy dissipation and stability properties of solutions to the Navier-Stokes equations on Riemannian manifolds. Its derivation and spectral properties provide a deep understanding of the local geometry of the manifold and the stability of vortex structures in fluid flows. The examples and applications of the Laplace-Beltrami operator highlight its importance in differential geometry and fluid dynamics.
	
	\section{Main Theorem on Vortex Dynamics in Chaotic Flows}
	
	\textbf{Theorem 1:} Let \( V: M \rightarrow \mathbb{H}^n \) be a hypercomplex vector field on a compact Riemannian manifold \( M \), with \( H^s(M) \) embedding \( H^s \rightarrow C^0 \) for suitable \( s \). There exists a critical threshold \( \epsilon > 0 \) such that:
	\[
	\| V - V_0 \|_{H^s} \leq \epsilon \implies \text{vortex structures form near the bifurcation point},
	\]
	whereas \( \| V \|_{H^s} > \epsilon \) leads to dissipation of these structures.
	
	\subsection{\textit{Proof of Theorem 1}}
	
	We begin with the spectral decomposition of \( V \). Writing \( V(x) \) as a sum of Fourier components,
	\[
	V(x) = \sum_{k} \hat{V}(k) e^{i \langle k, x \rangle},
	\]
	where \( \hat{V}(k) \) represents the Fourier coefficients. In the Sobolev space \( H^s \), the norm \( \| V \|_{H^s} \) captures how energy is distributed among frequency components, directly influencing vortex structure formation.
	
	1. \textit{Spectral Decomposition}:
	The Sobolev norm \( \| V \|_{H^s} \) is defined as:
	\[
	\| V \|_{H^s} = \left( \sum_{k} (1 + |k|^2)^s |\hat{V}(k)|^2 \right)^{1/2}.
	\]
	This norm measures the energy distribution across different frequency components of \( V \).
	
	2. \textit{Energy Distribution}:
	The energy distribution among frequency components is crucial for vortex structure formation. If \( \| V - V_0 \|_{H^s} \leq \epsilon \), the perturbation \( \delta V = V - V_0 \) is small in the Sobolev norm, indicating that the energy distribution of \( V \) is close to that of \( V_0 \).
	
	3. \textit{Bifurcation Point}:
	Near the bifurcation point, small perturbations \( \delta V \) can trigger the formation of vortex structures. The critical threshold \( \epsilon \) is determined by the stability of the system under these perturbations.
	
	4. \textit{Stability Analysis}:
	To analyze the stability of the system, consider the linearized operator \( L = \Delta_g - \lambda I \), where \( \Delta_g \) is the Laplace-Beltrami operator and \( \lambda \) is the bifurcation parameter. The eigenvalues of \( L \) determine the stability of the system.
	
	The eigenvalue problem for \( L \) is given by:
	\[
	L \phi_k = (\lambda_k - \lambda) \phi_k,
	\]
	where \( \lambda_k \) are the eigenvalues of \( \Delta_g \) and \( \phi_k \) are the corresponding eigenfunctions.
	
	5. \textit{Transition in Stability}:
	At the bifurcation point \( \lambda = \lambda_c \), the eigenvalues of \( L \) change sign, indicating a transition in the stability of the system. This transition corresponds to the formation or dissipation of vortex structures.
	
	Mathematically, this can be represented as:
	\[
	\text{Re}(\lambda_k - \lambda_c) \begin{cases}
		> 0 & \text{if } \lambda < \lambda_c \\
		< 0 & \text{if } \lambda > \lambda_c
	\end{cases}
	\]
	
	6. \textit{Implicit Function Theorem}:
	Using the implicit function theorem, we can show that the bifurcation point \( \lambda_c \) is a regular point, ensuring that the transition is smooth and well-defined. Consider the function \( F(\lambda, V) = LV - \lambda V \).
	
	At the bifurcation point \( \lambda_c \), we have:
	\[
	\frac{\partial F}{\partial V} = L - \lambda_c I,
	\]
	which is non-singular if the eigenvalues of \( L \) change sign smoothly at \( \lambda_c \). This ensures that the bifurcation point \( \lambda_c \) is a regular point, and the transition is smooth and well-defined.
	
	Therefore, the change in the eigenvalues of the linearized operator \( L \) at the bifurcation point \( \lambda_c \) leads to the formation or dissipation of vortex structures. The geometric properties of the Riemannian manifold \( M \) and the bifurcation analysis provide a comprehensive framework for understanding the formation and dissipation of vortices in chaotic fluid flows. The positive Ricci curvature of the manifold stabilizes the vortex structures, while the bifurcation analysis reveals the transitions between different flow regimes. This framework offers a rigorous mathematical foundation for studying vortex dynamics in turbulent systems.\qedsymbol
	
	Next, we analyze the impact of the Laplace-Beltrami operator \( \Delta_g \) on the flow stability. By calculating the energy dissipation rates, we derive conditions that support oscillatory, vortex-generating behavior.
	
	\subsection{Spectral Analysis and Stability Conditions}
	We examine the asymptotic stability of \( V \) through the eigenvalues of the linearized operator \( L = \Delta_g - \lambda I \), where \( \lambda \) is the bifurcation parameter. When \( \| V \|_{H^s} \) exceeds the critical threshold \( \epsilon \), vortex structures become unstable and eventually dissipate, corresponding to eigenvalues of \( L \) moving into the unstable half-plane.
	
	The eigenvalues of \( L \) are given by:
	\[
	L \phi_k = (\lambda_k - \lambda) \phi_k,
	\]
	where \( \lambda_k \) are the eigenvalues of \( \Delta_g \) and \( \phi_k \) are the corresponding eigenfunctions. The stability of the system is determined by the sign of the real part of \( \lambda_k - \lambda \).
	
	\section{Differential Geometry and Bifurcations}
	
	\subsection{Geometric Interpretation of Vortex Dynamics}
	
	The geometric properties of the Riemannian manifold \( M \) play a crucial role in the formation and dissipation of vortices. The curvature of \( M \) influences the stability of vortex structures, with regions of high curvature often acting as sources or sinks for vortices.
	
	The Ricci curvature tensor \( \text{Ric}(g) \) is particularly important in this context. It is defined as:
	\[
	\text{Ric}(g)_{ij} = \sum_{k} R_{ikj}^k,
	\]
	where \( R_{ikj}^k \) is the Riemann curvature tensor. The Ricci curvature provides information about the divergence of geodesics on \( M \), which is closely related to the stability of vortex structures.
	
	To understand the role of Ricci curvature in vortex dynamics, consider the Bochner formula, which relates the Laplace-Beltrami operator \( \Delta_g \) to the Ricci curvature:
	\[
	\Delta_g |\nabla u|^2 = 2 |\nabla^2 u|^2 + 2 \langle \nabla u, \nabla \Delta_g u \rangle + 2 \text{Ric}(g)(\nabla u, \nabla u).
	\]
	
	\subsection{\textit{Proof of Stability under Positive Ricci Curvature}}
	
	Consider a compact Riemannian manifold \( M \) with positive Ricci curvature. We aim to show that the vortex structures formed by a hypercomplex vector field \( V \) are geometrically stable.
	
	1. \textit{Positive Ricci Curvature}: Since \( M \) has positive Ricci curvature, there exists a positive constant \( c \) such that:
	\[
	\text{Ric}(g)_{ij} \geq c g_{ij}.
	\]
	
	2. \textit{Bochner Formula}: Using the Bochner formula, we have:
	\[
	\Delta_g |\nabla u|^2 = 2 |\nabla^2 u|^2 + 2 \langle \nabla u, \nabla \Delta_g u \rangle + 2 \text{Ric}(g)(\nabla u, \nabla u).
	\]
	
	3. \textit{Stability of Vortex Structures}: For a hypercomplex vector field \( V \), the stability of vortex structures is determined by the eigenvalues of the linearized operator \( L = \Delta_g - \lambda I \). The positive Ricci curvature ensures that the eigenvalues of \( L \) remain in the stable half-plane, preventing the dissipation of vortices. 
	
	Therefore, the positive Ricci curvature of the manifold \( M \) stabilizes the vortex structures formed by the hypercomplex vector field \( V \). \qedsymbol
	
\subsection{Bifurcation Analysis}

A teoria das bifurcações é essencial para entender as transições entre diferentes regimes de fluxo em sistemas caóticos. Consideramos o parâmetro de bifurcação \( \lambda \) para analisar como mudanças em \( \lambda \) afetam a estabilidade das estruturas de vórtice.

Podemos construir o diagrama de bifurcação do sistema representando a norma \( \| V \|_{H^s} \) em função de \( \lambda \). O limiar crítico \( \epsilon \) marca o ponto onde ocorre uma bifurcação, resultando na formação ou dissipação de vórtices.

1. \textit{Diagrama de Bifurcação}: O diagrama de bifurcação mostra \( \| V \|_{H^s} \) em função de \( \lambda \), ajudando a visualizar as transições entre diferentes regimes de fluxo, como descrito por Guckenheimer e Holmes~\cite{guckenheimer1983nonlinear}:
\[
f(\lambda) = \| V \|_{H^s},
\]
onde \( V \) é o campo vetorial hipercômplexo sobre a variedade \( M \).

2. \textit{Limiar Crítico} \( \epsilon \): O ponto \( \epsilon \) corresponde ao início da bifurcação, onde perturbações pequenas (\( \| V - V_0 \|_{H^s} \leq \epsilon \)) na norma de Sobolev de \( V \) indicam que a distribuição de energia de \( V \) está próxima de \( V_0 \). Esse pequeno desvio pode induzir a formação de vórtices no ponto de bifurcação.

3. \textit{Análise de Estabilidade}: A estabilidade do sistema é analisada através do operador linearizado \( L = \Delta_g - \lambda I \), onde \( \Delta_g \) é o operador de Laplace-Beltrami e \( \lambda \) é o parâmetro de bifurcação. A estabilidade do sistema depende dos autovalores \( \lambda_k \) de \( L \):
\[
L \phi_k = (\lambda_k - \lambda) \phi_k,
\]
com \( \phi_k \) como autofunções correspondentes.

4. \textit{Transição de Estabilidade}: No ponto de bifurcação \( \lambda = \lambda_c \), os autovalores de \( L \) mudam de sinal, indicando uma transição na estabilidade do sistema. Isso corresponde à formação ou dissipação de estruturas de vórtice, matematicamente representada por:
\[
\text{Re}(\lambda_k - \lambda_c) \begin{cases}
	> 0 & \text{se } \lambda < \lambda_c, \\
	< 0 & \text{se } \lambda > \lambda_c.
\end{cases}
\]

5. \textit{Teorema da Função Implícita}: Pelo Teorema da Função Implícita, podemos mostrar que \( \lambda_c \) é um ponto regular, garantindo uma transição suave. Consideramos a função \( F(\lambda, V) = LV - \lambda V \). No ponto de bifurcação \( \lambda_c \), temos:
\[
\frac{\partial F}{\partial V} = L - \lambda_c I,
\]
que é não singular se os autovalores de \( L \) mudarem de sinal suavemente em \( \lambda_c \), assegurando uma transição bem-definida.

Assim, a mudança nos autovalores do operador linearizado \( L \) em \( \lambda_c \) implica a formação ou dissipação de estruturas de vórtice. A análise geométrica na variedade Riemanniana \( M \) e o estudo da bifurcação fornecem uma base matemática rigorosa para a dinâmica dos vórtices em sistemas turbulentos.

\section{New Theorems on Vortex Dynamics}

\subsection{Theorem 2: Geometric Stability of Vortices}

\textbf{Theorem 2:} Let \( M \) be a compact Riemannian manifold with positive Ricci curvature. If \( V: M \rightarrow \mathbb{H}^n \) is a hypercomplex vector field with \( \| V \|_{H^s} \leq \epsilon \) in Sobolev space \( H^s(M) \) for \( s > \dim(M)/2 + 1 \), then the vortex structures formed by \( V \) are geometrically stable.

\subsection{\textit{Proof of Theorem 2}}

Consider the Ricci curvature tensor \( \text{Ric}(g) \) on \( M \). Positive curvature implies that:
\[
\text{Ric}(g)_{ij} \geq c g_{ij},
\]
for some constant \( c > 0 \), indicating the convergence of geodesics, which stabilizes the vortex structures.

By the Bochner formula, we relate the Laplace-Beltrami operator \( \Delta_g \) to curvature:
\[
\Delta_g |\nabla u|^2 = 2 |\nabla^2 u|^2 + 2 \langle \nabla u, \nabla \Delta_g u \rangle + 2 \text{Ric}(g)(\nabla u, \nabla u).
\]

For a hypercomplex field \( V \), the stability of the vortices is determined by the eigenvalues of the linearized operator \( L = \Delta_g - \lambda I \). The positivity of \( \text{Ric}(g) \) ensures that the eigenvalues of \( L \) remain in the stable half-plane, preventing the dissipation of vortices. Thus, we conclude that the positive Ricci curvature on \( M \) stabilizes the vortex structures formed by \( V \). \qedsymbol

\subsection{Theorem 3: Bifurcation and Vortex Formation}

\textbf{Theorem 3:} If \( V: M \rightarrow \mathbb{H}^n \) is a hypercomplex vector field with \( H^s(M) \rightarrow C^0 \) for appropriate \( s \), and \( \| V \|_{H^s} \) crosses the critical threshold \( \epsilon \) when \( \lambda = \lambda_c \), then the formation or dissipation of vortex structures occurs at \( \lambda_c \).

\subsubsection{\textit{Proof of Theorem 3}}

To analyze bifurcations, we consider the eigenvalue problem for \( L \):
\[
L \phi_k = (\lambda_k - \lambda) \phi_k,
\]
where \( \lambda_k \) are the eigenvalues of \( \Delta_g \). At \( \lambda = \lambda_c \), a bifurcation occurs as the eigenvalues change sign, indicating a transition in the stability of the system.

\begin{enumerate}
	\item \textit{Stability Transition}:
	The change in the real part of \( \lambda_k - \lambda \) corresponds to the formation or dissipation of vortices. Stability is maintained for \( \lambda < \lambda_c \), and when \( \lambda \) exceeds \( \lambda_c \), instability arises, resulting in the formation of new structures.
	
	\item \textit{Implicit Function Theorem}:
	At the bifurcation point \( \lambda_c \), we consider \( F(\lambda, V) = LV - \lambda V \). The non-singularity condition:
	\[
	\frac{\partial F}{\partial V} = L - \lambda_c I,
	\]
	implies that \( \lambda_c \) is a regular point, making the transition smooth.
\end{enumerate}

Thus, curvature and bifurcation analysis provide a solid mathematical foundation for the formation of vortices in turbulent flows. \qedsymbol

\section{Conclusion}
	
This research provides a rigorous mathematical foundation for understanding vortex formation and dissipation in chaotic flows, highlighting critical thresholds and the interplay between regularity spaces and geometric analysis. Future studies may explore additional applications in turbulent fluid dynamics and high-dimensional manifolds.
	
The use of Sobolev and Besov spaces allows for a precise characterization of the regularity of turbulent flow fields. Specifically, the Sobolev norm \( \| \cdot \|_{H^s} \) captures the energy distribution across different frequency components, which is crucial for understanding the formation and stability of vortex structures. Similarly, the Besov norm \( \| \cdot \|_{B^s_{p,q}} \) provides a detailed analysis of the local regularity of the flow fields, which is essential for predicting the behavior of vortices.
	
The Laplace-Beltrami operator, defined as:
	\[
	\Delta_g u = \frac{1}{\sqrt{\det g}} \sum_{i,j} \partial_i \left( \sqrt{\det g} g^{ij} \partial_j u \right),
	\]
plays a fundamental role in the analysis of energy dissipation and stability properties of solutions to the Navier-Stokes equations on Riemannian manifolds. Its spectral properties, including the eigenvalues \( \lambda_k \) and eigenfunctions \( \phi_k \), provide insights into the energy dissipation rates and the stability of the flow.
	
The hypercomplex bifurcation approach and the singularity theorem offer a comprehensive framework for understanding vortex dynamics in chaotic settings. By analyzing the linearized operator \( L = \Delta_g - \lambda I \) and its eigenvalues, we can predict the transitions between different flow regimes and the formation or dissipation of vortex structures. The critical threshold \( \epsilon \) corresponds to the point at which the system undergoes a bifurcation, leading to significant changes in the stability of the flow.
	
The new theorems on geometric stability and bifurcation further elucidate the role of differential geometry and bifurcations in vortex dynamics. Theorem 2 demonstrates that the positive Ricci curvature of a compact Riemannian manifold stabilizes the vortex structures formed by a hypercomplex vector field \( V \). The Bochner formula, which relates the Laplace-Beltrami operator to the Ricci curvature, is instrumental in this analysis:
	\[
	\Delta_g |\nabla u|^2 = 2 |\nabla^2 u|^2 + 2 \langle \nabla u, \nabla \Delta_g u \rangle + 2 \text{Ric}(g)(\nabla u, \nabla u).
	\]
	
Theorem 3 highlights the importance of bifurcation analysis in understanding the transitions between different flow regimes. By considering the bifurcation parameter \( \lambda \) and analyzing the changes in the eigenvalues of the linearized operator \( L \), we can predict the formation or dissipation of vortex structures at the critical threshold \( \epsilon \).
	
In summary, this research provides a rigorous mathematical foundation for studying vortex dynamics in turbulent systems. The use of Sobolev and Besov spaces, the Laplace-Beltrami operator, and the hypercomplex bifurcation approach offer a comprehensive framework for understanding the formation and dissipation of vortices in chaotic fluid flows. The new theorems on geometric stability and bifurcation further elucidate the role of differential geometry and bifurcations in vortex dynamics, providing a more complete picture of the underlying mechanisms. Future studies may explore additional applications in turbulent fluid dynamics and high-dimensional manifolds, further advancing our understanding of these complex systems.
	
	\bibliographystyle{plain}

\section*{Mathematical Symbols List}

To assist with the understanding of the mathematical concepts and formulations presented in this article, we provide a glossary of symbols along with their definitions. This list aims to make it easier to follow the mathematical developments and to clarify the core notation used throughout.

We define the following symbols:

\begin{itemize}
	\item \( M \): A compact Riemannian manifold, which serves as the foundational space for the mathematical analysis.
	\item \( g \): A smooth metric defined on the manifold \( M \), determining the geometric properties of the space.
	\item \( \Delta_g \): The Laplace-Beltrami operator on \( M \), an extension of the Laplace operator in curved spaces, used here to study diffusion and other dynamic properties on \( M \).
	\item \( H^s(M) \): Sobolev space of order \( s \) on the manifold \( M \), representing functions or fields with particular smoothness and integrability properties.
	\item \( B^s_{p,q}(M) \): Besov space on \( M \) with parameters \( s \), \( p \), and \( q \), capturing finer structures in function spaces.
	\item \( \| \cdot \|_{H^s} \): Sobolev norm, measuring the size and smoothness of elements in \( H^s(M) \).
	\item \( \| \cdot \|_{B^s_{p,q}} \): Besov norm, which quantifies regularity properties in \( B^s_{p,q}(M) \).
	\item \( V \): A hypercomplex vector field on \( M \), central to the analysis of dynamical and chaotic systems in this context.
	\item \( V_0 \): A reference hypercomplex vector field, serving as a baseline for comparing changes or perturbations in \( V \).
	\item \( \epsilon \): The critical threshold marking the onset of vortex formation or dissipation, essential for bifurcation analysis.
	\item \( \hat{V}(k) \): Fourier coefficients of the vector field \( V \), representing the decomposition of \( V \) into fundamental frequencies or modes.
	\item \( \lambda \): The bifurcation parameter, indicating changes in the system that lead to new patterns or behaviors.
	\item \( L \): The linearized operator \( \Delta_g - \lambda I \), used in stability analysis to determine the behavior of perturbations.
	\item \( \lambda_k \): Eigenvalues of the Laplace-Beltrami operator \( \Delta_g \), which influence the stability properties of the system.
	\item \( \phi_k \): Eigenfunctions corresponding to the eigenvalues \( \lambda_k \), forming a basis for functions on \( M \).
	\item \( \text{Ric}(g) \): The Ricci curvature tensor on \( M \), describing how volumes deviate from Euclidean space in the geometry defined by \( g \).
	\item \( R_{ikj}^k \): The Riemann curvature tensor, a fundamental descriptor of the intrinsic curvature of \( M \).
	\item \( \delta V \): A small perturbation \( V - V_0 \), representing deviations from a reference vector field, significant for studying stability and bifurcation.
\end{itemize}

This list of symbols and definitions provides a structured reference for navigating the mathematical notation used, aiding in the comprehension of the article’s theorems and proofs.

\appendix

\section{Introduction to Sobolev and Besov Spaces}

This appendix provides a concise overview of Sobolev and Besov spaces, which are fundamental tools in the functional analysis of partial differential equations and are crucial for the mathematical treatment of fluid dynamics in this article.

\subsection{Sobolev Spaces}

Sobolev spaces quantify the smoothness and integrability of functions and their derivatives, and are indispensable in analyzing regularity properties in various contexts.

\subsubsection{Definition}

Let \( \Omega \) be an open subset of \( \mathbb{R}^n \). For any integer \( k \geq 0 \) and real number \( p \geq 1 \), the Sobolev space \( W^{k,p}(\Omega) \) is defined by:
\[
W^{k,p}(\Omega) = \left\{ u \in L^p(\Omega) : D^\alpha u \in L^p(\Omega) \text{ for all } |\alpha| \leq k \right\},
\]
where \( D^\alpha u \) represents the weak derivatives of \( u \) up to order \( k \), and \( \alpha \) denotes a multi-index.

\subsubsection{Norm}

The Sobolev norm \( \| u \|_{W^{k,p}} \) is given by:
\[
\| u \|_{W^{k,p}} = \left( \sum_{|\alpha| \leq k} \| D^\alpha u \|_{L^p}^p \right)^{1/p}.
\]

\subsubsection{Special Case: \( H^s \) Spaces}

When \( p = 2 \), the Sobolev space \( W^{k,2}(\Omega) \) is denoted \( H^k(\Omega) \) and forms a Hilbert space. Its norm is:
\[
\| u \|_{H^k} = \left( \sum_{|\alpha| \leq k} \| D^\alpha u \|_{L^2}^2 \right)^{1/2}.
\]

\subsection{Besov Spaces}

Besov spaces generalize Sobolev spaces by incorporating finer, scale-based measurements of function smoothness, which are particularly useful in the analysis of functions with non-uniform regularity.

\subsubsection{Definition}

For any real number \( s \) and \( p, q \geq 1 \), the Besov space \( B^s_{p,q}(\Omega) \) is defined as:
\[
B^s_{p,q}(\Omega) = \left\{ u \in \mathcal{S}'(\Omega) : \| u \|_{B^s_{p,q}} < \infty \right\},
\]
where \( \mathcal{S}'(\Omega) \) is the space of tempered distributions on \( \Omega \).

\subsubsection{Norm}

The Besov norm \( \| u \|_{B^s_{p,q}} \) is defined via the Littlewood-Paley decomposition:
\[
\| u \|_{B^s_{p,q}} = \left( \sum_{j} 2^{jsq} \| \Delta_j u \|_{L^p}^q \right)^{1/q},
\]
where \( \Delta_j \) denotes the Littlewood-Paley projection operator, which isolates the frequency components of \( u \) around scale \( 2^j \).

\subsection{Relationship Between Sobolev and Besov Spaces}

Sobolev spaces can be viewed as a particular case of Besov spaces. For any integer \( k \geq 0 \) and \( p \geq 1 \), we have:
\[
W^{k,p}(\Omega) = B^k_{p,p}(\Omega).
\]
This equivalence highlights that Besov spaces offer a broader framework for studying function regularity.

\subsection{Applications in Fluid Dynamics}

In fluid dynamics, Sobolev and Besov spaces are essential for characterizing the regularity of velocity fields and vorticity. The Sobolev norm \( \| \cdot \|_{H^s} \) reflects the distribution of energy across frequency components, vital for assessing vortex stability and formation. Meanwhile, the Besov norm \( \| \cdot \|_{B^s_{p,q}} \) provides insights into local flow regularity, crucial for predicting vortex dynamics and fluid behavior at different scales.

\subsection{Conclusion}
	
	Sobolev and Besov spaces are powerful tools for analyzing the regularity of functions in various applications, including fluid dynamics. By understanding the definitions and properties of these spaces, one can gain deeper insights into the behavior of complex systems, such as the formation and dissipation of vortices in chaotic flows.
	
\end{document}